\documentclass[fleqn,usenatbib]{mnras}

\DeclareRobustCommand{\VAN}[3]{#2}
\let\VANthebibliography\thebibliography
\def\thebibliography{\DeclareRobustCommand{\VAN}[3]{##3}\VANthebibliography}

\usepackage{graphicx}	% Including figure files
\usepackage{amsmath}	% Advanced maths commands
\usepackage{amssymb}	% Extra maths symbols
\usepackage{xcolor}
\usepackage{soul}
\definecolor{seagreen}{rgb}{0.190, 0.725, 0.161}
\definecolor{red}{rgb}{0.545, 0., 0.}

\title[GW190521 formation via three-body encounters]{GW190521 formation via three-body encounters in young massive star clusters}

\author[Dall'Amico et al.]{
Marco Dall'Amico$^{1,2}$\thanks{E-mail: \href{mailto:marco.dallamico@pd.infn.it}{marco.dallamico@pd.infn.it}},
Michela Mapelli$^{1,2,3}$\thanks{E-mail: \href{mailto:michela.mapelli@unipd.it}{michela.mapelli@unipd.it}},
Ugo N. Di Carlo$^{1,2,3}$,
Yann Bouffanais$^{1,2}$,
\newauthor{Sara Rastello$^{1,2}$, Filippo Santoliquido$^{1,2}$, Alessandro Ballone$^{1,2,3}$,
Manuel Arca Sedda$^4$}
\\
$^{1}$Physics and Astronomy Department Galileo Galilei, University of Padova, Vicolo dell’Osservatorio 3, I–35122, Padova, Italy\\
$^{2}$INFN-Padova, Via Marzolo 8, I–35131 Padova, Italy\\
$^{3}$INAF–Osservatorio Astronomico di Padova, Vicolo dell’Osservatorio 5, I–35122, Padova, Italy\\
$^{4}$Astronomisches Rechen-Institut, Zentr\"um f\"ur Astronomie, Universit\"at Heidelberg, M\"onchofstr. 12-14, Heidelberg, Germany
}

\date{Accepted XXX. Received YYY; in original form ZZZ}

\pubyear{2021}

\begin{document}
\label{firstpage}
\pagerange{\pageref{firstpage}--\pageref{lastpage}}
\maketitle

\begin{abstract}
GW190521 is the most massive binary black hole (BBH) merger observed to date, and its primary component lies in the pair-instability (PI) mass gap. Here, we investigate the formation of GW190521-like systems via three-body encounters in young massive star clusters.  We performed 2$\times10^5$ simulations of binary-single interactions between a BBH and a massive $\geq{60}\,$M$_{\odot}$ black hole (BH), including post-Newtonian terms up to the $2.5$ order and a  prescription for relativistic kicks. In our initial conditions, we take into account the possibility of forming BHs in the PI mass gap via stellar collisions. If we assume that first-generation BHs have low spins, $\sim{0.17}\%$ of all the simulated BBH mergers have component masses, effective and precessing spin, and remnant mass and spin inside the $90\%$ credible intervals of GW190521. Seven of these systems are first-generation exchanged binaries, while five are second-generation BBHs. We estimate a merger rate density  $\mathcal{R}_{\rm GW190521}\sim{0.03}\,$Gpc$^{-3}\,$yr$^{-1}$ for GW190521-like binaries formed via binary-single interactions in young star clusters. This rate is extremely sensitive to the spin distribution of first-generation BBHs. Stellar collisions, second-generation mergers and dynamical exchanges are the key ingredients to produce GW190521-like systems in young star clusters. 
\end{abstract}


%%%%%%%%%%%%%%%%%%%%%%%%%%%%%%%%%%%%%%%%%%%%%%%%%%

%%%%%%%%%%%%%%%%% BODY OF PAPER %%%%%%%%%%%%%%%%%%

\section{Introduction}

Since the detection of GW150914 \citep{abbottGW150914,abbottGW150914astro}, the number of gravitational wave (GW) sources observed by the LIGO--Virgo collaboration (LVC) has increased year after year, culminating with the recent publication of the results of the first half of the third LVC observing run  %second gravitational wave transient catalog 
\citep{AbbottGWTC21,AbbottO3}. So far, the sample of detected compact binaries includes 53 binary black hole (BBH) candidates, 2 binary neutron stars \citep{Abbott2017,Abbott2019} and 2 possible neutron star -- black hole binary systems \citep{AbbottNSBH}. Among these systems, GW190521 detains the record of the most massive BBH ever observed, with primary mass $m_{1}=85^{+21}_{-14}\,$M$_{\odot}$ and secondary mass $m_{2}=66^{+17}_{-18}\,$M$_{\odot}$ in the source frame ($90\%$ credible interval, \citealt{AbbottGW190521,AbbottGW190521astro}). The coalescence of these two massive black holes (BHs) produced a $\sim{140}\,$M$_{\odot}$ remnant that lies in the still unexplored intermediate mass range of the BH mass spectrum, and can thus be considered as the first intermediate-mass BH (IMBH) candidate detected with GWs \citep{AbbottGW190521astro}. IMBHs bridge the gap between stellar-mass and super-massive BHs in the range $10^{2}\leq m_{\rm BH}/$M$_{\odot}\leq 10^{5}$; their existence is pivotal to explain the nature of ultra- and hyper-luminous X-ray sources and the growth of super-massive BHs \citep[e.g.,][for a recent review]{Greene2019}.

The primary BH of GW190521 has a $99\%$ probability of lying in the pair-instability (PI) mass gap ($\sim{60-120}\,$M$_{\odot}$, \citealt{AbbottGW190521,AbbottGW190521astro}, see also \citealt{Mehta2021}). In this mass range, no BH is expected to form from the collapse of a single star, as a consequence of the unstable oxygen-silicon burning phase experienced by the progenitor \citep{heger2002,woosley2007,belczynski2016b,spera2017,woosley2017,marchant2019,stevenson2019,woosley2019,woosley2021}. \cite{Fishbach2020} and \cite{Nitz2021} interpret GW190521 as a merger event straddling the PI mass gap. In this case, the primary mass would safely be above the upper edge of the mass gap.

 GW190521 shows mild evidence for precession effects \citep{AbbottGW190521}. The waveform analysis reports a precessing spin parameter  $\chi_{\mathrm{p}}=0.68^{+0.25}_{-0.37}$, and an effective spin parameter  $\chi_{\mathrm{eff}}=0.08^{+0.27}_{-0.36}$ ($90\%$ credible interval), favouring a precessing binary model with in-plane spin components and high spin magnitudes for both BHs. Finally, some authors also claim support for non-zero eccentricity at the time of merger (\citealt{gayathri2020,romero2020,bustillo2020}). 

Because of its peculiar properties, the formation of GW190521 is still a matter of debate. First, the boundaries of the PI mass gap still suffer from large uncertainties, mostly related to nuclear reaction rates, stellar rotation and the fate of the outer envelope (\citealt{farmer2019,farmer2020,Farrell2020,tanikawa2020,umeda2020,Mapelli2020b,renzo2020a,costa2020}). Assuming a recent estimate of such uncertainties, % on the PI mass gap, 
\cite{belczynski2020} shows that it is possible to produce a system with similar masses to GW190521 via isolated binary evolution. 
On the other hand, this scenario can hardly account for a strong spin misalignment. Binary evolution tends to align the spin of the two components with the orbital angular momentum vector, and the BHs that result from the direct collapse of the two stars inherit their spin orientation forming a non-precessing BBH \citep{gerosa2018,bavera2020}. In contrast, a dynamically active environment tends to isotropically redistribute the spin orientation of BHs \citep{rodriguez2016spin}, while also favouring the production of higher mass binaries \citep{hills1980}. 

In the hierarchical merger scenario, a BH can undergo repeated mergers with smaller BHs, as long as it is harbored in a star cluster \citep{miller2002}. This mechanism has recently been studied by several authors to explain the origin of GW190521 \citep{fragione2020,anagnostou2020,kimball2020,mapelli2021,arcasedda2021}. Hierarchical mergers can also take place inside the disc of active galactic nuclei, where BBHs such as GW190521 can form in the migration trap due to the dynamical friction exerted by the disc \citep{mckernan2012,mckernan2018,bartos2017,samsing2020,laszlo2020,secunda2020,tagawa2021a,tagawa2021b}. Either in the core of a dense star cluster or in the disc of a galactic nucleus, the coalescence of GW190521 might also have been caused by Kozai-Lidov oscillations \citep{kozai1962,lidov1962} induced on the binary by the central super-massive BH \citep{liu2021}. Finally, several authors \citep{spera2019,dicarlo2019,dicarlo2020a,dicarlo2020b,gerosa2021} have shown that PI-mass range BHs can be formed from the collapse of a massive star with an oversized hydrogen-rich envelope and a relatively small helium core. This object could be the product of single or repeated stellar collisions between stars with a well-developed helium core and main-sequence/Hertzsprung-gap stars  \citep{kremer2020,renzo2020b,gomez2021,gonzalez2021}. 

Here, we study the dynamical formation of systems like GW190521 in young star clusters (YSCs), by means of $2\times10^5$ three-body simulations with post-Newtonian terms. 
We start from the results of the simulations by \cite{dicarlo2019}. Since it would be computationally prohibitive to study a large sample of GW190521-like systems with full \textit{N}-body simulations, we extract the main properties of our single and binary BHs (mass and semi-major axis distribution) from the simulations by \cite{dicarlo2019} and we use them to simulate the formation of GW190521-like systems with three-body encounters. In this way, we include BHs in the mass gap formed via stellar mergers.

\section{Methods}\label{sec:methods}

\subsection{\textit{N}-body simulations with ARWV}
\label{sec:ARWV} 

We simulated $2\times10^{5}$ three-body encounters between a BBH and a single massive BH  using the direct \textit{N}-body code {\sc arwv} \citep[][]{arcasedda2019,chassonnery2019,chassonnery2021}. {\sc arwv} exploits the algorithmic regularization chain method to integrate the equations of motion  \citep{mikkola1989,mikkola1993}. For our simulations we make use of the {\sc arwv} feature to combine the logarithmic-Hamiltonian regularization (logH, \citealt{mikkola1999a,mikkola1999b,preto1999}) with the Time-Transformed-Leapfrog method (TTL, \citealt{mikkola2002}). The code implements a post-Newtonian (PN) treatment up to the 2.5 order for the correction of the equations of motion in case of strong gravitational interaction \citep{mikkola2008, memmesheimer2004}. {\sc arwv} calculates the relativistic kick received by the BH remnant due to anisotropic GW emission at merger adopting the equations reported by \cite{healy2018}.

We integrate each three-body encounter for $10^{5}$ yr. If at that time the system is still in an unstable triple configuration, the simulation is then restarted and carried on until the conclusion of the interaction. At the end of the simulation, if only a BBH is left, we calculate its merger time as \citep{peters1964}
\begin{eqnarray}\label{eq:peters}
 \frac{{\rm d}a}{{\rm d}t}=-\frac{64}{5}\,{} \frac{G^3 \,{} m_{i} \,{} m_{j} \,{} (m_{i}+m_{j})}{c^5 \,{} a^3\,{} (1-e^2)^{7/2}} \,{} f_1(e)\nonumber\\
  \frac{{\rm d}e}{{\rm d}t}=-\frac{304}{15}\,{} e \frac{ G^3 \,{} m_{i} \,{} m_{j} \,{} (m_{i}+m_{j})}{c^5 \,{}a^4 \,{}  (1-e^2)^{5/2}}\,{}f_2(e),
\end{eqnarray}
where $G$ is the gravity constant, $c$ the speed of light, $m_{i}$ the primary mass, $m_{j}$ the secondary mass, $a$ the semi-major axis, $e$ the orbital eccentricity and
\begin{eqnarray}
f_1(e)=\left(1+\frac{73}{24}\,{}e^2+\frac{37}{96}\,{} e^4\right) \nonumber\\
f_2(e)=\left(1+\frac{121}{304} \,{} e^2\right).
\end{eqnarray}
These equations only account for the effect of GW emission, they do not encode the information relative to the first and second post-Newtonian terms. For this reason, we integrate a system with equations \ref{eq:peters} only after the three-body interaction is concluded and only a binary is left. 

We assume that two BHs merge when their distance is $\leq{}6\,{}G(m_{i}+m_{j})/c^{2}$, i.e. the sum of the innermost stable circular orbits of the two BHs considering non-spinning BHs.

\subsection{Initial Conditions}\label{IC}

We set our three-body scattering experiments in the massive YSCs of \cite{dicarlo2019}. This family of clusters can be frequently found in star-forming spiral, starburst and interacting galaxies, including the Milky Way (e.g., see \citealt[][]{portegies2010} for a review). In YSCs, star formation is still at work, and they are one of the main forges of massive stars in the local Universe \citep[][]{lada2003}. Several studies in the literature have already shown that YSCs are ideal birthplaces for BBH mergers  \citep{portegies2000,portegies2002,portegies2010,ziosi2014,mapelli2016,kimpson2016,chatterjee2017b,banerjee2017,banerjee2018a,banerjee2018b,banerjee2020,dicarlo2019,dicarlo2020a,dicarlo2020b,kumamoto2019,kumamoto2020,trani2021}.

In the simulations of \cite{dicarlo2019} and \cite{dicarlo2020a}, massive stars rapidly sink toward the core of the star cluster, where they may experience repeated collisions with other massive stars and thus increase their mass. This can cause stars to acquire a large hydrogen-rich envelope maintaining a relatively small helium core ( $\lesssim{32}\,$M$_{\odot}$). If the star concludes its life before mass loss efficiently erodes its envelope, and its  core does not grow above the threshold for PI, the star avoids PI and instead collapses directly to form a BH in the $60-120\,$M$_{\odot}$ mass range. Since the direct collapse mechanism does not induce a strong recoil kick on the compact remnant, these BHs likely remain inside the YSC and can pair-up dynamically, possibly leading to the formation of BBHs \citep{heggie2003}.

Our sample of synthetic three-body simulations is generated considering YSCs with a metallicity  $Z=0.002\simeq{}0.1$ Z$_\odot$. Star clusters with lower metallicity develop BBH populations with a similar mass spectrum \cite[e.g.,][]{dicarlo2020b}. In contrast, at higher metallicity, the formation of BHs in the PI mass gap and IMBH mass range is quenched by wind mass loss episodes experienced by the stellar progenitors along their evolution \citep{dicarlo2020a}. 

From here on, we will refer to the quantities related to the primary BH with the subscript 1, to the secondary BH with 2, and to the single BH with 3. Moreover, to distinguish the initial configuration from the outcome binaries, we call \textit{original binary} and \textit{intruder} respectively the BBH ($m_{1}-m_{2}$) and the single BH ($m_{3}$) that are generated from the initial conditions and set as input to the simulation at time $t=0$.

The initial conditions for BH masses are extracted from the simulations of YSCs performed by \cite{dicarlo2019}, considering the most massive clusters of their sample with $8\times10^{3}\le M_{\rm cl}/{\rm M}_\odot\le{}3\times10^{4}$.  The simulations of \cite{dicarlo2019} implement realistic models for stellar and binary evolution and allow the formation of BHs in the PI mass gap and in the IMBH mass range via repeated stellar mergers. This is pivotal in our study since it allows our initial BHs to be a representative sample of the BH population of a YSC. 
We derive three independent distributions for $m_{1}$, $m_{2}$ and $m_{3}$ applying the kernel density estimation method to the BBHs and single BH populations of \cite{dicarlo2019}.
We then randomly sample the BH masses from these distributions. Based on \cite{dicarlo2019}, we draw the mass of primary BHs in the 
$[3.7, 438]\,{\rm M}_\odot$ range, the mass of secondary BHs in the $[3, 74]\,{\rm M}_{\odot}$ range and the mass of the intruder in the  $[60, 378]\,{\rm M}_{\odot}$. Namely, we specifically require the intruder mass to be above the lower end of the PI mass gap. The main reason for this choice is that BHs in the PI mass gap are only $\sim{}1$\% of the entire population by \cite{dicarlo2019}: if we had simulated intruders with all possible masses, including lower mass BHs, we would have needed to run $\sim{}100$ times more simulations, with a prohibitive computational cost. When calculating the merger rate of GW190521-like systems, we will correct our results accounting for the whole possible intruder mass range.

The semi-major axes $a$ are derived from the simulations of \cite{dicarlo2019} and \cite{dicarlo2020a}. We fit a log-normal distribution to their data with mean $\mu_{\log{(a/{\rm AU})}}=1.51$ and sigma $\sigma_{\log{(a/{\rm AU})}}=0.92$, and then we randomly sample from this distribution to generate the initial semi-major axis of our original BBHs. We set the limits of the distribution to $[5.8\times10^{-2}, 10^{4}]$ AU, where the lower limit refers to the smallest semi-major axis in the sample of \cite{dicarlo2019}, while the upper limit is a cut-off value that we introduce to exclude soft binaries, using a 3D velocity dispersion of $5$\,km\,s$^{-1}$ as reference value for YSCs.

YSCs have a relatively short two-body relaxation timescale of $t_{\rm rlx}\sim{}20$~Myr~$(M_{\rm cl}/10^4\,{}{\rm M}_\odot)^{1/2}\,{}(r_{\rm h}/1\,{}{\rm pc})^{3/2}$, where $r_{\rm h}$ is the half-mass radius \citep{spitzer1987}. If the cluster reaches two-body relaxation, the stellar velocities can be described with a Maxwell-Boltzmann distribution. We assume the BHs are in thermal equilibrium with the cluster population, and we randomly generate the velocity at infinity $v_{\infty}$ from a Maxwellian distribution with a 3D velocity dispersion $\sigma_{\infty}=5$ km~s$^{-1}$, typical of a YSC. The sampled velocity can be interpreted as the relative velocity between the intruder and the centre-of-mass of the original binary. If the BBHs are in thermal equilibrium with the rest of the cluster population, the eccentricity values $e$ of the original binaries follow the thermal eccentricity distribution \citep{Ambartsumian,heggie1975}. This is further confirmed by the simulations of \cite{dicarlo2019}: they find that the eccentricity distribution of their BBHs at $100$\,Myr is coherent with this behaviour.
We thus generate the eccentricities from a uniform distribution in $e^{2}$ inside the range $[0, 1)$.

The GW events observed so far seem to favour a slowly spinning BH population \citep[][]{AbbottO3,AbbottO3popandrate}. Therefore, we generate the initial dimensionless spin of each BH  $\chi_{i}=S_{i}\,{}c/(G\,{}m_{i}^{2})$, where $S_{i}$ is the magnitude of the spin vector,  % and  $\,{}i=1,\,{}2,\,{}3$, 
according to a Maxwell-Boltzmann distribution with root-mean square $\sigma_\chi{}=0.1$, as already done by \cite{bouffanais2019,bouffanais2021}.  Star cluster dynamics tends to isotropically redistribute the natal spin direction of the BHs via dynamical encounters, which cause BHs to lose memory of their initial spin orientation with respect to the orbital plane of the BBH. To account for this effect, we randomly draw the spin directions isotropic over the sphere. We also check the main effects of a different choice of the spin magnitudes ($\sigma_\chi=0.01,$ 0.2, 0.3, 0.5) by overriding the spin magnitudes a posteriori, without rerunning the dynamical simulations.

For the remaining initial quantities (the impact parameter, the three orientation angles, and the phase of the binary star), we use the same formalism as \cite{hut1983}. The orientation of the encounter is randomly drawn from an isotropic sphere\footnote{The angles $\phi$, $\psi$ and $\theta$ are defined as in \cite{hut1983}: $\phi{}$ is the angle between the pericentre of the binary orbit and the intersection of the vertical plane in which lies the initial velocity vector of the intruder; $\psi{}$ is the angle that defines the orientation of the impact parameter with respect to the orbital plane direction in a surface perpendicular to the initial velocity of the intruder; the angle $\theta{}$ defines the aperture included between the perpendicular versor of the binary orbital plane and the intruder initial velocity direction at infinity.}, where the angles $\phi{}$ and $\psi{}$ are sampled from a uniform distribution in $[0, 2\,{}\pi)$, while $\theta{}$ is uniformly drawn from $\cos{\theta{}}$ in $[-1, 1]$.

We set the initial single-binary distance $D=100\,a$, so that the original binary is initially unperturbed by the intruder. The impact parameter $b$ is drawn according to a uniform probability distribution in $b^{2}$, due to its proportionality to the surface element transverse to the incoming direction of the intruder. The values are generated in the interval $[0,b_{\rm max}]$, with the upper limit derived from the gravitational focusing expression \citep{sigurdsson1993}:
\begin{equation}\label{eq:phinney}
b_{\rm max}=\frac{\sqrt{\,{}2\,{}G\,{}(m_{1}+m_{2}+m_{3})\,{}a}}{v_{\infty}},
\end{equation}
which represents the maximum impact parameter for a hard encounter as a function of the mass of each BH and the semi-major axis $a$ of the original BBH. Per each scattering experiment, we require that $b<D$. Equation \ref{eq:phinney} assumes that the simulated interactions have maximum pericentre $p_{\rm max}$ equal to the semi-major axis of the binary system. Three-body interactions with $p_{\rm max}>a$ likely lead to soft encounters, in which the energy exchange between the single body and the binary is negligible, and the system configuration remains unchanged. This implies that all our simulations are hard encounters. Including softer encounters in our simulations would have requested a larger number of runs, with a much higher computational cost.

The orbital phase of the original binary $f$ is generated in the range $[-\pi, \pi]$ according to the prescription adopted by \citet{hut1983}. For each original binary, we first derive the eccentric anomaly $\epsilon$ from
\begin{equation}\label{eq:anom}
\mathcal{F}=\epsilon{}-e\,{}\sin{\epsilon{}}.
\end{equation}
In equation~\ref{eq:anom}, $\mathcal{F}\equiv{}\frac{2\,{}\pi}{T}\,{}t_{\rm p}$, where $T$ is the orbital period of the BBH and $t_{\rm p}$ is the time elapsed since pericentre passage. We randomly sample $\mathcal{F}$ in the range $[0, 2\,{}\pi{})$. Finally, we retrieve the initial value of the binary phase $f$ with: 
\begin{equation}
\tan{\left(\frac{f}{2}\right)}=\left(\frac{1+e}{1-e}\right)^{1/2}\tan{\left(\frac{\epsilon{}}{2}\right)}.
\end{equation}

\section{Results}\label{sec:results}

\subsection{Flybys, exchanges and ionizations}

Three-body encounters are chaotic dynamical interactions that can evolve into several stable configurations. The outcome of an encounter strictly depends on the amount of energy exchanged in the process. 
In our simulations, we consider three possible outcomes: flybys, exchanges and ionizations. With \textit{flyby} we refer to any simulation in which the initial and final configuration of the three BHs is conserved, i.e. where the original binary survives to the three-body encounter. In this process, the binary can acquire binding energy (hardening) or lose it (softening), according to Heggie's law \citep[][]{heggie1975}. We define \textit{ionizations} all the events in which the binding energy of each BH pair is positive at the end of the simulation.
An ionization can happen only if the intruder approaches the binary with a velocity higher than the critical one \citep[][]{hut1983}:
\begin{equation}\label{eq:hut}
v_{\rm c} = \sqrt{\frac{\,{}G\,{} m_{1}\,{} m_{2}\,{}(m_{1}+m_{2}+m_{3})}{a\,{}m_{3}\,{} (m_{1}+m_{2})}}.
\end{equation}
If, at the end of the simulation, the resulting binary system is composed of different BHs with respect to the original ones, the encounter is labeled as an \textit{exchange} and the binary is an \textit{exchanged binary}. Exchange events are the product of resonant or prompt interactions during which the intruder replaces the primary or secondary BH of the original BBH to form an exchanged binary. The probability of an exchange to happen is higher if the intruder is more massive than one of the two binary members \citep{hills1980}. Thus, the final exchanged binary tends to have a higher total mass than the initial one. Flybys and exchanges may induce two of the three BHs to merge during the simulation.  If the binding energy between the remnant BH and the third  BH is sufficiently large that the relativistic kick does not unbind the binary system, the remnant BH and the third BH form a new BBH, which, in turn, can merge again. We refer to these latter systems as \textit{second-generation} BBHs. In contrast, if the remaining binary after the interaction does not contain a BH remnant (i.e., after an exchange or a fly-by event) it is defined as a \textit{first-generation} BBH.

Table~\ref{tab:out} reports the outcome fractions of our three-body experiments. Overall, the flybys represent $\approx{}18$\% of all the simulations, while exchanges are the most common outcome  ($\approx{}79$\%). The BBH is ionized only in the $\approx{}3$\% of the simulations. Table~\ref{tab:out2} focuses on the BBH mergers (i.e., all the simulated BBHs that merge within a Hubble time). Over a total of $7187$ BBH mergers, $54\%$ ($25.5$\%) are exchanged binaries where the secondary (primary) component is kicked off the system, $20.2\%$ are flybys and $0.3\%$ are second-generation BBHs. 

We calculate the merger timescale (eq.~\ref{eq:peters}) at the beginning of all the simulations ($\tau_0$) using the initial orbital properties of the BBHs. After the three-body simulation, we calculate again the merger timescale adopting the new orbital properties of the BBH ($\tau_{\rm 1g}$), and we define this timescale as the time-span between the beginning of the three-body integration and the merger. The values of  $\tau_{\rm 1g}$ and $\tau_0$ can be different because of the perturbations induced by the three-body encounter, which might speed up or delay the merger. In $0.25\%$ of the simulations, we observe the merger of the first-generation BBH during the three-body simulation. About $91\%$ of the BBHs that merge during the three-body integration (i.e., $\tau_{\rm 1g}<10^5$ yr) have an initial delay time of $\tau_0>10^5\,$yr: their coalescence is sped up by the three-body encounter.  Finally, in $0.005\%$ of the simulations, we have a second-generation BBH merger during a timescale $\tau_{\rm 2g}$, defined as the time elapsed from the beginning of the simulation.

%%%%%%%%%%%%%%%%%%%%%%%%%%%%%%%%%%%%%%%%%%%TABLE%%%%%%%%%%%%%%%%%%%%%%%%%%%%%%%%%%%%%%%%
\begin{table}
	\begin{center}
	\caption{Outcomes of three-body encounters for all the simulations.}
	\label{tab:out}
	\begin{tabular}{lccc} 
		\hline
		$\mathit{f}_{\mathrm{fb}}$  & $\mathit{f}_{\mathrm{ex13}}$  & $\mathit{f}_{\mathrm{ex23}}$  & $\mathit{f}_{\mathrm{ion}}$  \\
        \hline
        
        0.178 & 0.521 & 0.274 & 0.027\\
        
		\hline
	\end{tabular}
	\end{center}
	\flushleft
	\footnotesize{Column 1 ($f_{\rm fb}$): fraction of flybys; column 2 ($f_{\rm ex13}$): fraction of exchanges in which the final BBH is composed of $m_1$ and $m_3$ (the secondary BH was kicked off); column 3 ($f_{\rm ex23}$): fraction of exchanges in which the final BBH is composed of $m_2$ and $m_3$ (the primary BH was kicked off); column 4 ($f_{\rm ion}$): ionization fraction. }
\end{table}
%%%%%%%%%%%%%%%%%%%%%%%%%%%%%%%%%%%%%%%%%%%%%%%%%%%%%%%%%%%%%%%%%%%%%%%%%%%%%%%%%%%%%%%%%

%%%%%%%%%%%%%%%%%%%%%%%%%%%%%%%%%%%%%%%%%%%TABLE%%%%%%%%%%%%%%%%%%%%%%%%%%%%%%%%%%%%%%%%
\begin{table}
	\begin{center}
	\caption{Outcomes of three-body encounters for BBH mergers only (first line) and of BBH mergers that match the masses of GW190521 (second line).}
	\label{tab:out2}
	\begin{tabular}{lcccc} 
		\hline
		     Sample                      &
             $\mathit{f}_{\mathrm{2G}}$  &  $\mathit{f}_{\mathrm{12}}$  & $\mathit{f}_{\mathrm{13}}$  & $\mathit{f}_{\mathrm{23}}$   \\
        \hline
        
        BBH mergers & 0.002 & 0.193 & 0.567 & 0.238 \\
        GW190521 & 0.006 & 0.009 & 0.971 & 0.014 \\
        
		\hline
	\end{tabular}
	\end{center}
		\flushleft
\footnotesize{Column 1: the considered sample can be all BBH mergers (first line) or only the mergers with component masses inside the 90\% credible interval of GW190521 (second line) according to \cite{AbbottGW190521}. Column 2 ($f_{\rm 2G}$): fraction of second-generation mergers (i.e., the merger remnant of the BBH merges with the third BH); column 3 ($f_{\rm 12}$): fraction of mergers between $m_1$ and $m_2$; column 4 ($f_{\rm 13}$): fraction of mergers between $m_1$ and $m_3$; column 5 ($f_{\rm 23}$): fraction of mergers between $m_2$ and $m_3$. }
\end{table}
%%%%%%%%%%%%%%%%%%%%%%%%%%%%%%%%%%%%%%%%%%%%%%%%%%%%%%%%%%%%%%%%%%%%%%%%%%%%%%%%%%%%%%%%%

%%%%%%%%%%%%%%%%%%%%%%%%%%%%%%%%%%%%%%%FIGURE 1%%%%%%%%%%%%%%%%%%%%%%%%%%%%%%%%
\begin{figure}
	\includegraphics[width=\columnwidth]{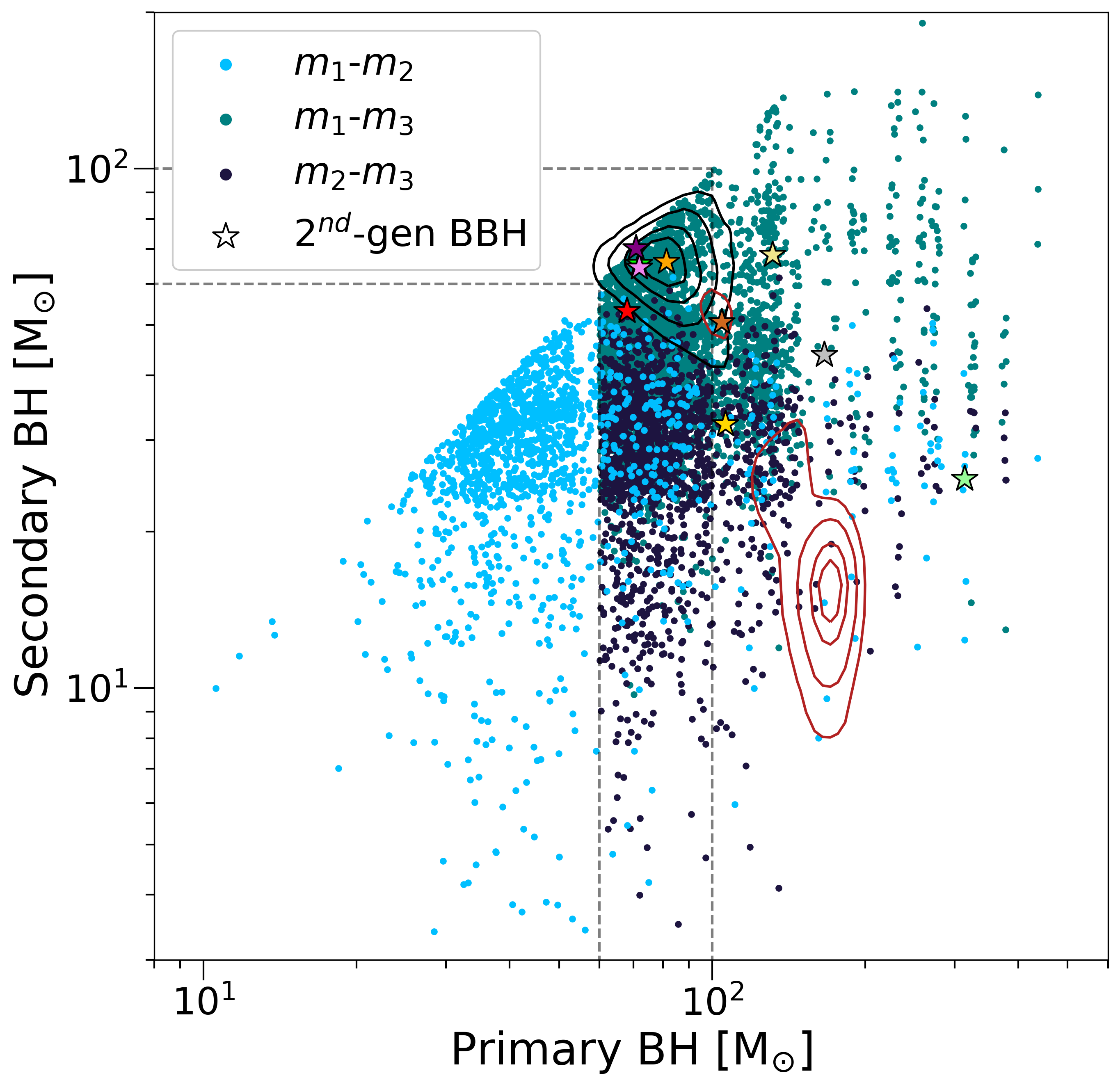}
    \caption{Primary and secondary masses of the simulated BBH mergers. Light blue circles are flyby BBHs, while grey (dark) blue circles are exchanged BBHs where the intruder replaced the secondary (primary) BH.  The black (magenta) contour levels are the 25, 50, 75, 90$\%$ credible regions of GW190521  reported by \protect{\citealt{AbbottGW190521}}  (\protect{\citealt{Nitz2021}}). Coloured stars are second-generation BBHs. The lime-green, brown, purple, orange and violet stars are inside the 90\% credible regions from \protect{\citealt{AbbottGW190521}}. The vertical dashed grey lines mark the lower-end of the PI mass gap, at $60$ M$_{\odot}$, and the lower end of the IMBH mass range, at $100$ M$_{\odot}$.}
    \label{fig:mass_scatter}
\end{figure}
%%%%%%%%%%%%%%%%%%%%%%%%%%%%%%%%%%%%%%%%%%%%%%%%%%%%%%%%%%%%%%%%%%%%%%%%%%%%%%%%

%%%%%%%%%%%%%%%%%%%%%%%%%%%% FIGURE 2 %%%%%%%%%%%%%%%%%%%%%%%%%%%%%%%%%%%%%%%%%%
\begin{figure}
	\includegraphics[width=\columnwidth]{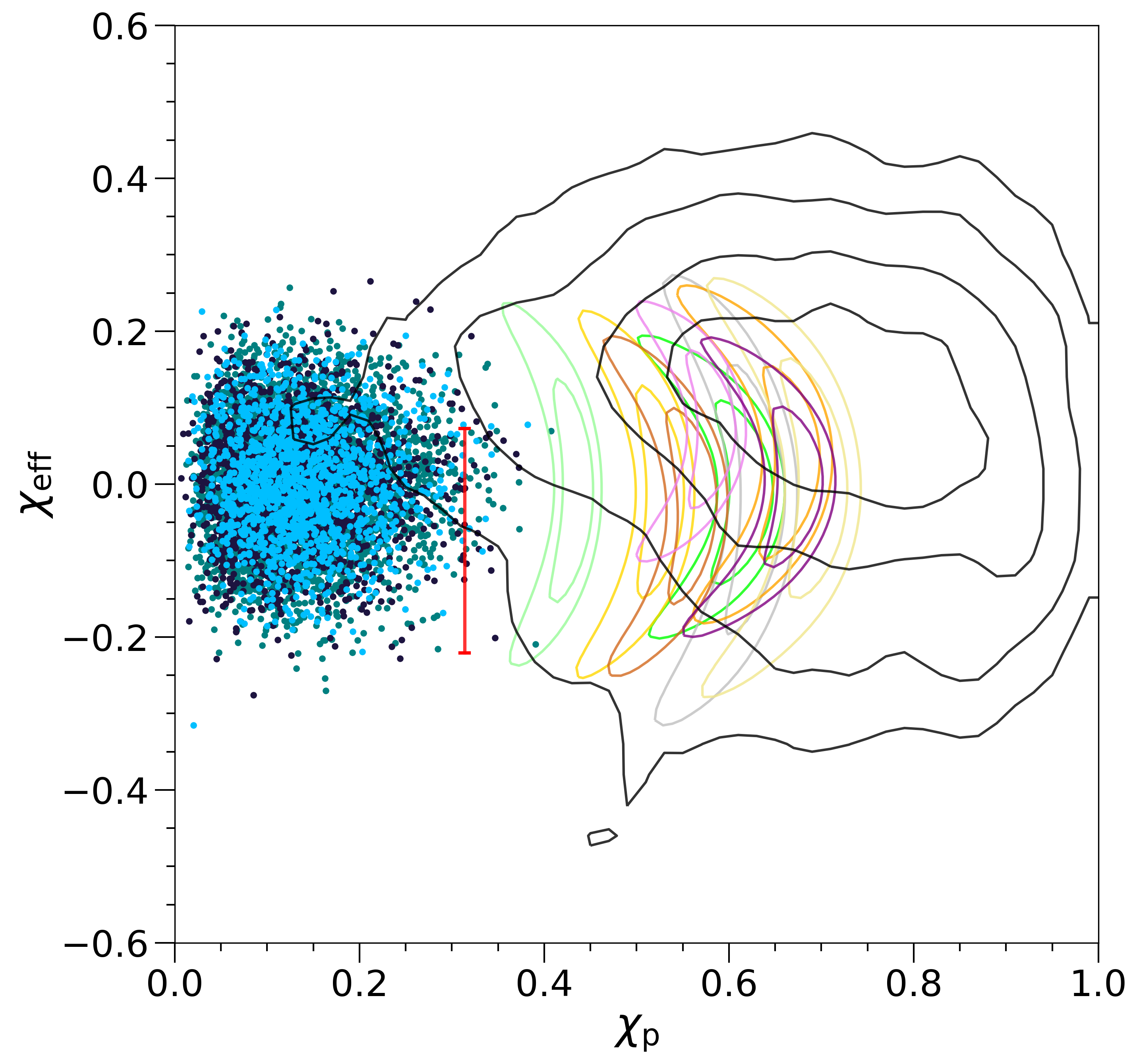}
    \caption{ Effective spin parameter $\chi_{\rm eff}$ versus precessing spin parameter $\chi_{\rm p}$ for all the BBH mergers. The colours are the same as Figure~\ref{fig:mass_scatter}. The lime-green, gray, orange, yellow, brown, light-green, violet, khaki, purple contours are the $50$ and $90\%$ credible regions for 9 out of the 10 second-generation BBHs. The red bar shows the last second-generation BBH for which $\chi_{\rm p}$ depends only on the spin of the first-generation component (see the main text for details). The black contours are the 25, 50, 75, 90$\%$ credible regions for the GW190521 spin parameters posterior reported by \protect{\citealt{AbbottGW190521}} and \protect{\citealt{AbbottGW190521astro}}.
}
    \label{fig:spins_scatter}
\end{figure}
%%%%%%%%%%%%%%%%%%%%%%%%%%%%%%%%%%%%%%%%%%%%%%%%%%%%%%%%%%%%%%%%%%%%%%%%%%%%%%%%%

\subsection{Component masses}

Figure \ref{fig:mass_scatter} shows the mass of the primary and secondary components of the BBH mergers.
We now focus only on the BBH mergers that have both the primary and secondary mass in the $90\%$ credible intervals of GW190521 ($85^{+21}_{-14}$ and $66^{+17}_{-18}$ M$_\odot$, as reported by \citealt{AbbottGW190521,AbbottGW190521astro}). 
One every $\sim{}9$ BBH mergers ($11\%$ of the total) satisfy this criterion.  As shown by Table~\ref{tab:out2}, the vast majority of these systems are exchanged BBHs ($98.5$\%). Most of these mergers are between $m_1$ and $m_3$ ($97.1$\%), while mergers between $m_2$ and $m_3$ are only the $1.4\%$ of the GW190521-like systems. Flybys and second-generation binaries contribute to $0.9\%$ and  $0.6\%$ of the GW190521-like systems, respectively. Specifically, five over 10 second-generation BBHs lie inside the \cite{AbbottGW190521}  $90\%$ credible regions for the  component masses of GW190521. Their properties are reported in Table \ref{tab:gw190521-like}. In four of these three-body simulations, the original binary experiences a strong encounter with the intruder BH, during which $m_{3}$ extracts enough internal energy from the binary to induce it to merge. Despite the relativistic kick, the merger remnant resulting from this first coalescence forms a second-generation BBH with the intruder BH. These systems merge again in less than a Hubble time. The coalescence time of the original binary $m_1-m_2$ computed at the beginning of the simulation is longer than the duration of the simulation (i.e., $10^{5}\,$yr) for all of these mergers, meaning that the coalescence between $m_1$ and $m_2$ is sped up by the three-body interaction. One out of five second-generation BBHs matching the component masses of GW190521 is instead the product of an exchange event. In this simulation, the primary BH $m_1$ is kicked out from the original binary by the intruder, which merges with the secondary BH giving rise to a massive remnant. The remnant and $m_1$, in turn, form a second-generation BBH that merges again in less than a Hubble time.
Finally, another second-generation binary grazes the 90\% contours, but lies outside the 90\% credible interval of GW190521.

\cite{Nitz2021} interpret the detection of GW190521 as the coalescence of a BBH with  primary mass $168^{+15}_{-61}$ M$_\odot$ and  secondary mass $16^{+33}_{-3}$ M$_\odot$, according to the $90\%$ credible intervals derived with a uniform in mass-ratio prior \citep[see also][]{Fishbach2020,ezquiaga2021}. Their posterior distributions for the component masses are less populated by our BBH mergers than the posterior credible region of \cite{AbbottGW190521} and \cite{AbbottGW190521astro}. This may suggest that three-body encounters in YSCs could more easily create a BBH with both components in the $60-100\,$M$_{\odot}$ range than a binary straddling the PI mass gap. Indeed, even if a BBH is able to merge within the cluster, the BH remnant is likely kicked out by the relativistic recoil and cannot  participate to the cluster dynamics anymore. Hence, only a dense stellar environment with an escape velocity high enough to retain multiple-generation mergers \citep[e.g., nuclear star clusters; ][]{arcasedda2018,arcasedda2020b,antonini2019,fragione2020,fragione2021,mapelli2020,mapelli2021} is able to form an intermediate-mass ratio inspiral such as the one proposed by \cite{Nitz2021}.

\subsection{Effective and precessing spins}

Figure~\ref{fig:spins_scatter} shows the effective spin parameter $\chi_{\mathrm{eff}}$ as function of the precessing spin parameter $\chi_{\mathrm{p}}$ for all the BBH mergers of Figure~\ref{fig:mass_scatter}. These quantities are computed with the following expressions:
\begin{eqnarray}\label{eq:xeff}
\chi_{\rm eff}= \frac{(m_{i}\,{}\vec{\chi}_{i}+m_{j}\,{}\vec{\chi}_{j})}{m_{i}+m_{j}}\cdot{}\frac{\vec{L}}{L},
\nonumber{}\\
\chi_{\rm p}=\frac{c}{B_{i}\,{}G\,{}m_{i}^2}\,{}\max{(B_{i}\,{}S_{i\perp{}},\,{}B_{j}\,{}S_{j\,{}\perp})},
\end{eqnarray}
where $\vec{L}$ is the orbital angular momentum vector of the system, $S_{i\perp{}}$ and $S_{j\perp{}}$ are the spin angular momentum components in the orbital plane of the primary and secondary bodies of the binary, $B_i\equiv{}2+3\,{}q/2$ and $B_j\equiv{}2+3/(2\,{}q)$ with $q=m_{j}/m_{i}$ ($m_{i}\geq{}m_{j}$). Since dynamics randomly re-distributes the initial BH spins' orientation during a three-body interaction, we compute the final spin parameters $\chi_{\rm p}-\chi_{\rm eff}$ re-drawing the direction of each BH spin isotropically over a sphere but conserving their initial magnitude. For the BH remnants that pair up in second-generation BBHs we do not derive a single value but rather generate a full set of direction angles still sampled from an isotropic distribution. This implies that second-generation BBHs are represented in the plot as contour regions, with the  exception of one system (red bar) in which the first-generation component has a higher spin magnitude than the second-generation companion, and thus dominates the $\chi_{\rm p}$ term in equation~\ref{eq:xeff} resulting in one single $\chi_{\rm p}$ value for a set of $\chi_{\rm eff}$ values.

Figure~\ref{fig:spins_scatter} highlights two distinct populations of mergers. First-generation BBHs, which underwent exchanges and flybys, cover the parameter space at low values of the precessing spin, while second-generation BBHs are located at high $\chi_{\rm p}$. Half of all second-generation BBH mergers (five out of ten BBHs) match both the component masses and the spin parameters of GW190521 inside the 90\% credible regions reported by \cite{AbbottGW190521}, while only $0.1\%$ of the first-generation BBH mergers have both component masses and spin parameters inside the $90\%$ credible regions of GW190521 according to \cite{AbbottGW190521}. This is an effect of our assumption that all first-generation BH's spin magnitudes are distributed according to a Maxwellian distribution with $\sigma{}_\chi{}=0.1$. Had we assumed a larger value for $\sigma{}_{\chi}$, we would have obtained a correspondingly higher fraction of first-generation BBHs matching GW190521's component masses and spin parameters, as reported in Table \ref{tab:sigma_chi}.\\ 

The intersection of the two BBH samples that lie inside the posterior regions for the component masses (Figure~\ref{fig:mass_scatter}) and spin parameters (Figure~\ref{fig:spins_scatter}) of GW190521 contains twelve systems. These are five second-generation BBHs (marked by the lime-green, brown, purple, orange and violet stars in Figures \ref{fig:mass_scatter} and~\ref{fig:spins_scatter}) and seven exchanged binaries where $m_{3}$ replaced $m_{2}$ in the original system. The merger product of all these systems is an IMBH with a mass and a dimensionless spin magnitude inside the $90\%$ credible region of GW190521 ($M_{\rm rem}=142^{+28}_{-16}$ M$_\odot$ and $\chi_{\rm rem}=0.72^{+0.09}_{-0.12}$, \citealt{AbbottGW190521,AbbottGW190521astro}). Table~\ref{tab:gw190521-like} reports properties of the six BBHs matching GW190521, including the values of $\tau_0$, $\tau{}_{\rm 1g}$ and $\tau_{\rm 2g}$.

%%%%%%%%%%%%%%%%%%%%%%%%%%%%%%%%%%%%%%%%%%TABLE%%%%%%%%%%%%%%%%%%%%%%%%%%%%%%%%%%%%%%%%%%%%%
\begin{table*}
	\begin{center}
	\caption{BBH mergers with masses and spins in the 90\% credible intervals of GW190521. Upper five (Lower seven) lines: second-generation BBHs (first-generation BBHs).}
	\label{tab:gw190521-like}
	\begin{tabular}{lcccccccc} 
		\hline
        Name 2g & $m_i,\,{}m_j\,$[M$_{\odot}$] & $m_{\mathrm{rem}}^{\rm 1gen}-m_{k}\,$[M$_{\odot}$] & $m_{\mathrm{rem}}^{\rm 2gen}\,$[M$_{\odot}$] & $\chi_{\rm rem}^{\rm 2gen}$ & $\tau_{0}\,$[yr] & $\tau_{\mathrm{1g}}\,$[yr] & $\tau_{\mathrm{2g}}\,$[yr] & $e$\\ 
        \hline
        
        9721 & $35.4,\,{}33.3$   & $65.5-71.8$ & 130.7 & 0.67 & $1.1\times10^{5}$ & 1.52 & $4.1\times10^{4}$ & $2.6\times10^{-3}$\\
        19852 & $32.4,\,75.8$ & $104.3-50.6$  & 149.0 & 0.64 & $4.1\times10{5}$ & 1.0 & $6.13\times10^{8}$ & $9.0\times10^{-7}$\\
        86653 & $37.2,\,36.9$ & $70.5-70.8$ & 135.0 & 0.64 & $3.1\times10^{9}$ & 3.35 & $8.2\times10^{8}$ & $1.8\times10^{-6}$\\
        112964 & $43.3,\,{}42.0$ & $81.1-66.2$ & 139.9 & 0.70 & $2.7\times10^{5}$ & 0.79 & $6.5\times10^{4}$ & $4.0\times10^{-3}$\\
        128151 & $42.1,\,{}25.2$ & $64.3-71.9$ & 129.4 & 0.69 & $3.1\times10^{8}$ & 14.9 & $2.9\times10^{9}$ & $6.0\times10^{-7}$\\
		\hline
	Name 1g & $m_1,\,{}m_3\,$[M$_{\odot}$] & $m_{2}\,$[M$_{\odot}$] & $m_{\mathrm{rem}}\,$[M$_{\odot}$] & $\chi_{\rm rem}$ & $\tau_0$ [yr]& $\tau_{\mathrm{1g}}\,$[yr] & $\tau_{\mathrm{2g}}\,$[yr] & $e$\\
	\hline
	90086 & $73.7,\,{}70.8$ & 30.1 & 137.2 & 0.71 & $2.2\times10^{11}$ & $1.1\times10^{9}$ & -- & $1.0\times10^{-6}$ \\
	102042 & $78.2,\,{}85.7$ & 27.9 & 156.4 & 0.66 & $1.2\times10^{15}$ & $5.7\times10^{8}$  & -- & $1.6\times10^{-6}$\\
	129317 & $62.5,\,{}77.9$ & 30.1 & 134.0 & 0.66 & $1.1\times10^{11}$ & $8.4\times10^{5}$  & -- & $1.5\times10^{-5}$\\
	141880 & $52.0,\,{}84.7$ & 9.7  & 130.6 & 0.67 & $3.5\times10^{11}$ & $4.8\times10^{8}$  & -- & $1.3\times10^{-6}$\\
	154193 & $68.4,\,{}74.2$ & 6.9 & 135.5 & 0.71 & $1.6\times10^{7}$ & $8.3\times10^{9}$ & -- & $2.7\times10^{-7}$\\
	184181 & $48.1,\,{}68.7$ & 24.7 & 129.5 & 0.62 & $4.0\times10^{10}$ & $3.8\times10^{9}$ & -- & $2.4\times10^{-7}$\\
	188838 & $52.1,\,{}89.0$ & 47.2 & 134.9 & 0.68 & $9.2\times10^{13}$ & $1.4\times10^{9}$ & -- & $6.4\times10^{-7}$\\
	\hline
	\end{tabular}
	\end{center}
		\flushleft
\footnotesize{ The simulations in the first five lines are second-generation BBHs, reported in Figures~\ref{fig:mass_scatter}, \ref{fig:spins_scatter} and \ref{fig:mrem} with the colours lime-green, brown, purple, orange and violet. Subscripts $i$,$j$ in the first column and $k$ in the second column mark the three-body configuration that triggers the first merger: simulations 9721, 86653, 112964 and 128151 have $i=1,\,j=2$ (flyby) and $k=3$, while simulation 19852 has $i=2,\,j=3$ (exchange) and $k=1$. Column~1: simulation name;  column~2: mass of the components of the initial BBH; column~3: mass of the components of the second-generation BBH; column~4: mean mass of the final second-generation remnant BH; column~5: mean magnitude of the remnant spin; column~6 ($\tau_0$): initial coalescence time of the original BBH at the beginning of the simulation (we calculated the merger timescale for the initial BBH according to \protect{\citealt{peters1964}} assuming that the BBH is not perturbed by dynamics); column~7 ($\tau_{\rm 1g}$): effective coalescence time of the original binary as a result of the 3-body simulation; column~8 ($\tau_{\rm 2g}$): coalescence time of the second-generation BBH since the beginning of the simulation; column~9 ($e$): eccentricity at $\nu_{\rm gw}=10\,$Hz of the second-generation BBH.  The last seven lines are exchanged first-generation BBHs that match the properties of GW190521. Column~1: simulation name; column~2: mass of the components of the exchanged binary  (which is always composed of $m_1$ and $m_3$); column~3: mass of the secondary BH ejected during the exchange; column~4: mass of the final BH remnant; column~5: magnitude of the remnant spin; column~6: coalescence time of the BBH merger since the beginning of the simulation, calculated according to \protect{\cite{peters1964}}; column~7 ($\tau_{\rm 1g}$): effective coalescence time of the original binary as a result of the 3-body simulation; column~9 ($e$): eccentricity at $\nu_{\rm gw}=10\,$Hz of the exchanged BBH.}
\end{table*}
%%%%%%%%%%%%%%%%%%%%%%%%%%%%%%%%%%%%%%%%%%%%%%%%%%%%%%%%%%%%%%%%%%%%%%%%%%%%%%

%%%%%%%%%%%%%%%%%%%%%%
\begin{table}
	\begin{center}
	\caption{Percentage of first-generation BBH mergers that match the main properties of GW190521 as a function of the spin prescription adopted.}
	\label{tab:sigma_chi}
	\begin{tabular}{lc}
	    \hline
	    $\sigma_{\chi}$ & $\mathit{P}_{\rm GW190521}$ [\%]\\
		\hline
		0.01 & 0\\
		0.1 & 0.1\\
		0.2 & 2.8\\
		0.3 & 3.9\\
		0.5 & 4.2\\
	    \hline
	\end{tabular}
	\end{center}
		\flushleft
\footnotesize{Column~1 ($\sigma_{\chi}$): root-mean square value of the Maxwell-Boltzmann distribution used to generate the dimensionless spin magnitude of each BH. Column~2 ($\mathit{P}_{\rm GW190521}$): percentage of first-generation BBH mergers that have $m_1$, $m_2$, $\chi_{\rm eff}$, $\chi_{\rm p}$, $M_{\rm rem}$ and $\chi_{\rm rem}$ inside the $90\%$ credible intervals of GW190521 reported by \cite{AbbottGW190521,AbbottGW190521astro}.}
\end{table}

\subsection{Merger remnants}\label{subsec:properties}

%%%%%%%%%%%%%%%%%%%%%%%%%%%%%%%%%%%%%%%%%FIGURE 4%%%%%%%%%%%%%%%%%%%%%%%%%%%%%%%%%%%%%%%%%%%%
\begin{figure}
	\includegraphics[width=\columnwidth]{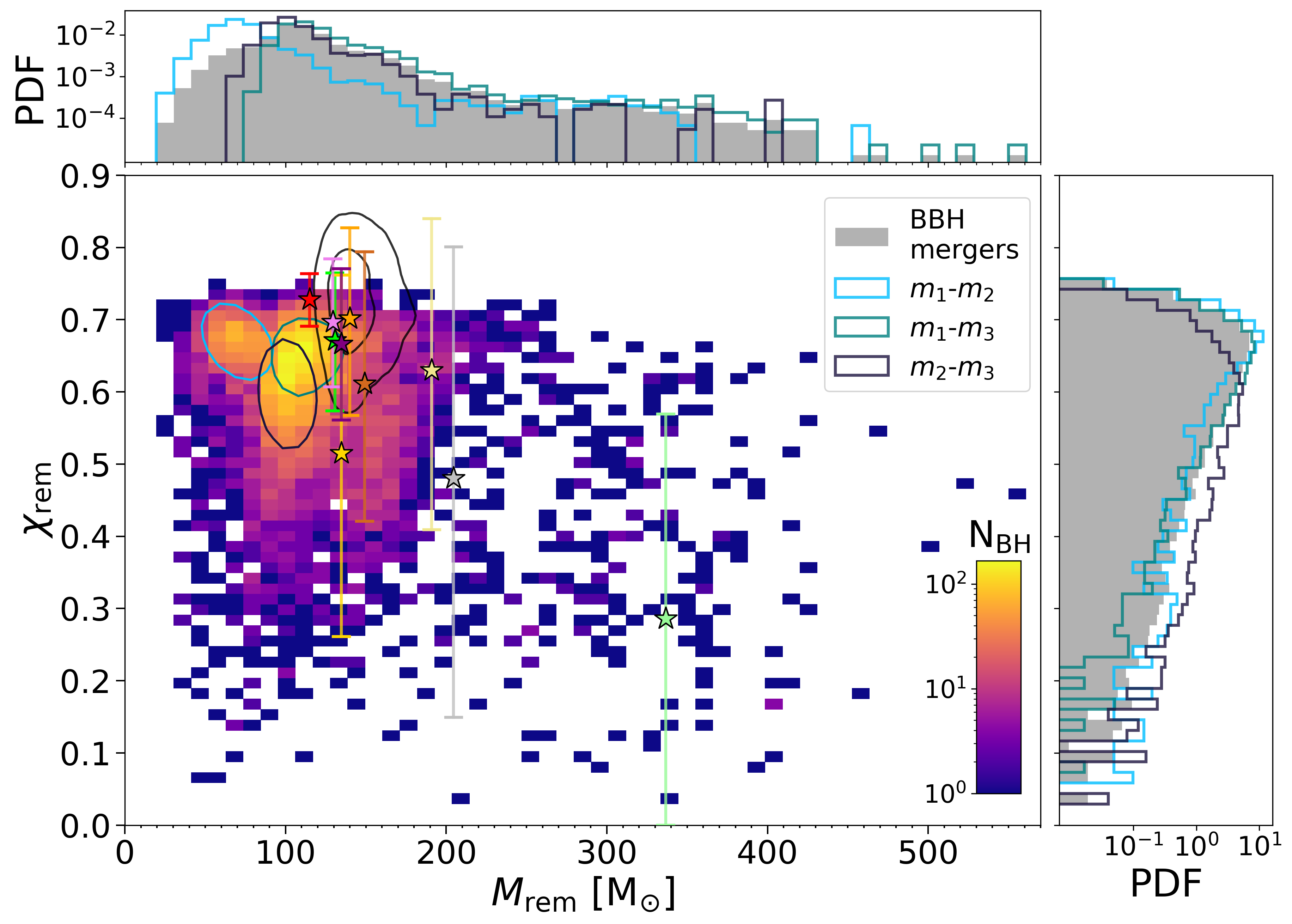}
    \caption{Mass of the BH remnant produced by each BBH merger as function of its dimensionless spin magnitude. The two-dimensional filled histogram shows all first-generation BBH mergers. The light-blue, dark-blue and navy unfilled contours show the $50\%$ credible regions for first-generation BBH megers with components $m_1-m_2$, $m_2-m_3$ and $m_1-m_3$, respectively. The stars mark the average values of $M_{\rm rem}$ and $\chi_{\rm rem}$ for second-generation BBH mergers, while the error bars show all the possible masses and spins inherited by these third-generation remnants (mass error bars are smaller than the markers, see the text for more details).
    The black unfilled contours show the $50$ and $90\%$ credible region for the posteriors of GW190521 \protect{\citep[][]{AbbottGW190521,AbbottGW190521astro}}. 
    The values of  of $M_{\rm rem}$ and $\chi_{\rm rem}$ for the lime-green, brown, purple, orange, violet and yellow stars are inside the 90\% credible region of GW190521.    
    The grey filled marginal histograms show the distributions of $M_{\rm rem}$ and $\chi_{\rm rem}$ for all simulated BBH mergers. The light-blue, dark-blue and navy unfilled marginal histograms show the distributions of  $M_{\rm rem}$ and $\chi_{\rm rem}$ for first-generation BBHs with  components $m_1-m_2$, $m_2-m_3$ and $m_1-m_3$, respectively.}
    \label{fig:mrem}
\end{figure}
%%%%%%%%%%%%%%%%%%%%%%%%%%%%%%%%%%%%%%%%%%%%%%%%%%%%%%%%%%%%%%%%%%%%%%%%%%%%%%%%%%%%%%%%%%

Figure \ref{fig:mrem} shows the mass of the merger remnants as function of their dimensionless spin magnitudes. The values are derived from the numerical relativity fitting equations of \cite{jimenez2017}. Specifically, to compute the remnant spin of the first-generation BBH mergers, we re-sampled the spin orientation of the progenitor BHs from an isotropic sphere. For second-generation BBHs we applied the same procedure we adopted in Figure \ref{fig:spins_scatter}: we randomly generated a full set of isotropic-oriented spins for the progenitor BHs, from which we then derived all the possible mass and spin magnitudes of the remnants. The error bars on the plot show all the possible masses and spins inherited by these third-generation remnants while the stars mark the mean value of the intervals.

The merger remnants inherit the orbital angular momentum of their progenitor BBH, and are therefore characterized by
high spin magnitudes \citep[][]{fishbach2017,gerosa2017}. The main peak of the distribution is located at a mass of $M_{\rm rem}\approx{}112\,$M$_{\odot}$ and a spin of $\chi_{\rm rem}\approx{}0.66$, and is mainly produced by exchanged BBHs with components $m_1-m_3$. Other two secondary peaks exist at $\chi_{\rm rem}\approx{}0.68$, $M_{\rm rem}\approx{}68\,$M$_{\odot}$ and $\chi_{\rm rem}\approx{}0.60$, $M_{\rm rem}\approx{}101\,$M$_{\odot}$, and are mostly given by the contribution of flybys $m_1-m_2$ and exchanged BBHs with components $m_2-m_3$, respectively. 

\begin{figure}
	\includegraphics[width=\columnwidth]{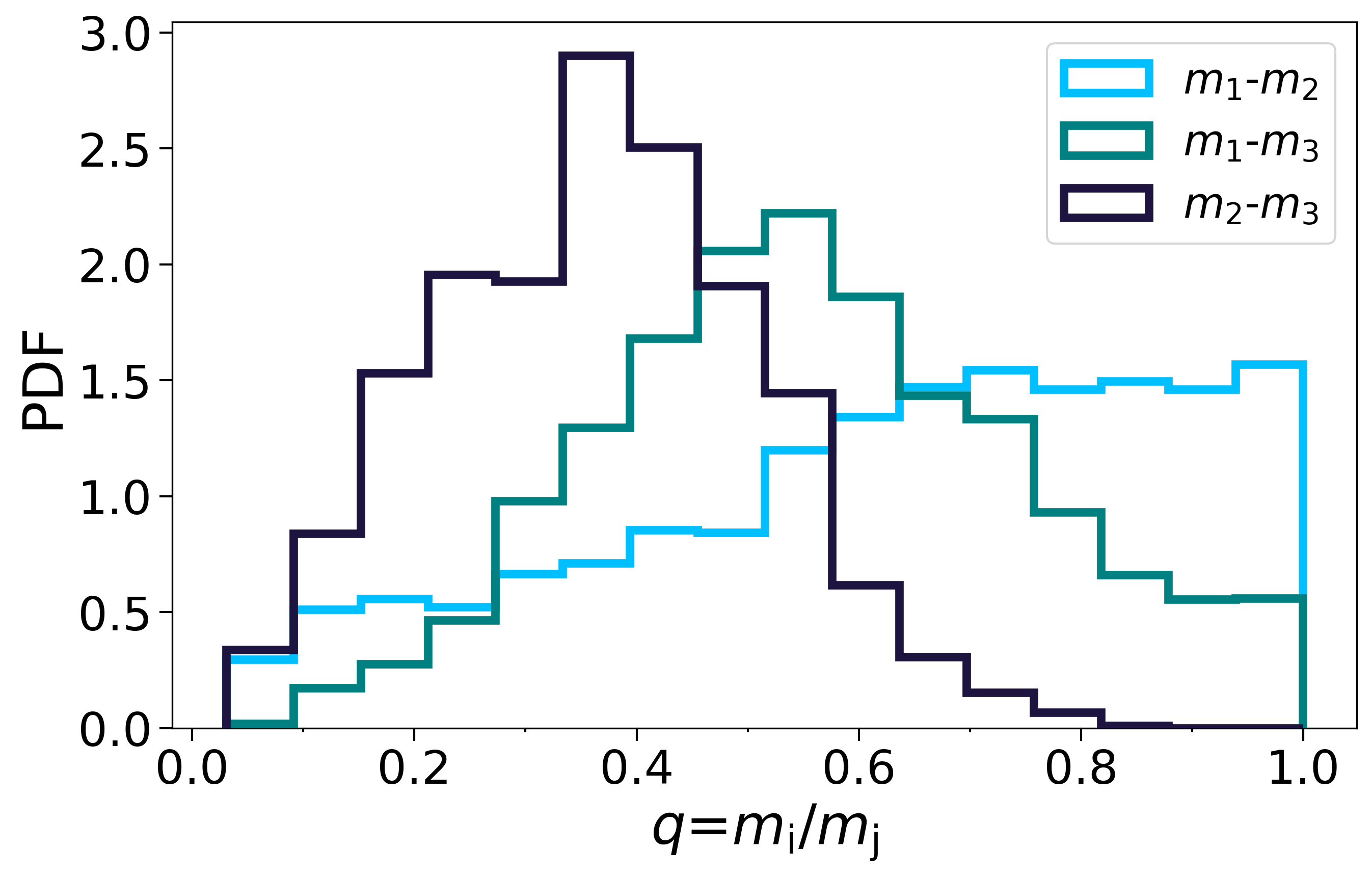}
    \caption{Mass ratio of the BBH mergers at the end of the simulations. The  histograms show the distribution of the three different outcomes: flybys are indicated with the light-blue line, while the navy (dark) blue lines show the exchanged binaries in which the intruder replaced the secondary (primary) BH.}
    \label{fig:mratio}
\end{figure}

The difference among these three sub-peaks is explained by the different total mass of the progenitor BBHs: flybys produce lower mass remnants than exchanged binaries, since the intruder ($m_3$) is generally more massive than the two members of the original BBH $m_1-m_2$. In their turn, exchanged binaries with component masses $m_1-m_3$ are more massive than exchanged binaries with component masses $m_2-m_3$, because $m_1>m_2$.
This difference in the BH masses results in a difference in the remnant spin $\chi_{\rm rem}$, mostly because of the different mass ratios. BBH mergers with components $m_1-m_2$, $m_1-m_3$ and $m_2-m_3$ have, on average,  different mass ratios with typical values of $\approx{}0.96$, $0.55$ and $0.36$, respectively (Figure \ref{fig:mratio}).

The only contour region that intersects the posteriors of GW190521 is the one populated mostly by exchanged BBHs with components $m_1-m_3$. Moreover, the same five second-generation BBHs that match the component masses of GW190521 lie inside the $90\%$ credible region of $M_{\rm rem}-\chi_{\rm rem}$, along with one additional system. This result further confirms that GW190521 might have been originated either by a primary exchange system or by a second-generation BBH.

%%%%%%%%%%%%%%%%%%%%%%%%%%%%%%%%%%%%%%%FIGURE 5%%%%%%%%%%%%%%%%%%%%%%%%%%%%%%%%%%%%%%%%%%%%%%%%%%
%%%%%%%%%%%%%%%%%%%%%%%%%%%%%%%%%%%%%%%%%%%%%%%%%%%%%%%%%%%%%%%%%%%%%%%%%%%%%%%%%%%%%%%%%%%%%%%%%

\section{Discussion}\label{sec:discussion}

\subsection{Dynamical origin of GW190521 and merger rate density}\label{sec:4.1}

There are at least four main features that characterize a BBH born from dynamical interactions with respect to an isolated binary.

\begin{itemize}
\item \textbf{Large total mass:} hierarchical BH mergers and repeated stellar collisions may produce massive BHs also inside the PI gap or even in the IMBH range  \citep[e.g.,][]{antonini2019,fragione2020,arcasedda2020a,mapelli2021}. These BHs can eventually interact with other binaries and form BBHs with higher total mass via exchange events.
\item \textbf{Misaligned spins:} dynamical interactions tend to isotropically redistribute the spin orientation of the binary components, while binary evolution in the isolated channel favours the production of parallel spinning stars due to angular momentum transfer \citep[e.g.,][]{rodriguez2016spin,gerosa2018}.
\item \textbf{Low mass ratio:} binary evolution can cause several stable and unstable mass transfer episodes, which generally redistribute the mass between the two stars and lead to a mass ratio close to one \citep[e.g.,][]{dominik2012,mapelli2019,neijssel2019}. On the other hand, in a dynamically active environment, exchange interactions produce lower mass ratio BBHs \citep[e.g.,][]{chatterjee2017a,dicarlo2019}.
\item \textbf{Non-zero eccentricity in chirping regime:} dynamical interactions and resonant perturbations such as the Kozai-Lidov effect can heavily increase the eccentricity in already existing BBHs, or even produce head-on collisions \citep[e.g][]{samsing2014,samsing2018,arcasedda2018,zevin2019}. These systems may merge before the GW emission is able to circularize the orbit, producing a distinct feature in the waveform \citep[e.g.,][]{gayathri2020,romero2020,holgado2021}. 
\end{itemize}

Our simulations indicate that GW190521 can be the result of a first-generation exchanged BBH with at least one component produced by a stellar merger, or of a second-generation BBH.
The posterior distribution of its component masses, the mass of the remnant, and the combination of its $\chi_{\rm p}-\chi_{\rm eff}$ spin parameters seem to exclude the merger of an original binary but rather favour a scenario in which a less massive BBH experienced an exchange event between the secondary and the massive intruder that increased the total mass of the system. Another interpretation is provided by the merger of a second-generation BBH. If the first-generation BH population is characterized by low spin magnitudes as suggested by \cite{AbbottO3}, the latter scenario is even more likely because of the mild evidence for large spins in GW190521 \citep{AbbottGW190521}. 

Some authors \citep{gayathri2020,romero2020,bustillo2020,AbbottGW190521astro} interpret the detection of GW190521 as the merger of a binary system with non-zero eccentricity at the time of coalescence. We calculated the eccentricity of the simulated BBH systems when the frequency of GW emission is $\nu_{\rm gw}=10$ Hz (i.e., approximately when the binary system enters the LIGO--Virgo range, Table~\ref{tab:gw190521-like}). We find that two of the second-generation BBHs that match the properties of GW190521 have respectively $e\sim{}0.003$ and $e\sim{}0.004$ in the LIGO--Virgo range (see table \ref{tab:gw190521-like}). This translates into $e\sim{0.4}$ and $e\sim0.3$ at $\nu_{\rm gw}=10^{-2}$ Hz in LISA band. All the other systems that match the properties of GW190521 have eccentricity $\lesssim10^{-4}$ in the LIGO--Virgo range, even if post-Newtonian corrections are accounted for.

Finally, we estimated the approximate merger rate density of GW190521-like systems from our simulations as
\begin{eqnarray}\label{eq:eq11}
    \mathcal{R}_{\rm GW190521}\sim{0.03}\,{\rm Gpc}^{-3}\,{\rm yr}^{-1}\left(\frac{N_{\rm 190521}}{12}\right)\,{}\left(\frac{N_{\rm BBH}}{7187}\right)^{-1}\nonumber\\
    \,{}\left(\frac{\mathcal{R}_{\rm BBH}(z=0.8)}{170\,{\rm Gpc}^{-3}\,{}{\rm yr}^{-1}}\right)\,\left(\frac{f_{\rm YSC}}{0.7}\right)\,\left(\frac{f_{\rm corr}}{0.14}\right),
\end{eqnarray}
 where $N_{\rm 190521}$ is the number of simulated BBH mergers with the mass of the components, the effective and precessing spin parameters and the mass and spin of the remnant inside the $90\%$ credible intervals reported by \cite{AbbottGW190521} and \cite{AbbottGW190521astro}, $N_{\rm BBH}$ is the number of BBH mergers in our simulations, $\mathcal{R}_{\rm BBH}(z=0.8)$ is the BBH merger rate density at $z\simeq{}0.8$ \citep[i.e., the median redshift value of GW190521; ][]{AbbottGW190521,AbbottGW190521astro}. We calculated $\mathcal{R}_{\rm BBH}$ for the YSCs simulated by \cite{dicarlo2020b} following the method described in \cite{santoliquido2020}. $\mathcal{R}_{\rm BBH}$ is affected by a substantial uncertainty (about one order of magnitude), mostly because of the metallicity evolution (see \citealt{santoliquido2021} for more details). Finally, $f_{\rm YSC}$ is the fraction of BBH mergers that originate in YSCs, according to the fiducial model of \cite{bouffanais2021}, and $f_{\rm corr}$ is a correction factor to compensate for the bias we introduced when we simulated only intruders with $m_3\geq{}60$ M$_\odot$. 
In the simulations of \cite{dicarlo2020a}, the BHs with mass inside the PI gap are only $\sim{}1\%$ of the whole BH population (considering both single and binary BHs), but the BBHs that contain at least one BH in the PI mass gap are $\sim{10\%}$ of all the BBHs. Since all BHs in the PI gap are single BHs at birth, this means that they are extremely efficient in pairing up via dynamical exchanges. In our three-body simulations, we find that $71\%$ of all the final BBHs have at least one component in the PI mass gap. Hence $f_{\rm corr}=0.14$ compensates for this spurious enhancement of BBHs in the mass gap in our simulations with respect to the ones of \cite{dicarlo2020a}.

Equation \ref{eq:eq11} leads to a merger rate density value of $\mathcal{R}_{\rm GW190521}\sim{}0.03\,$Gpc$^{-3}\,$yr$^{-1}$ for BBHs like GW190521 formed via three-body encounters in YSCs. This is about a factor of 2.7 lower than the median value reported in \cite{AbbottO3IMBH}, but still  inside their 90\% credible interval ($0.08^{+0.19}_{-0.07}$ Gpc$^{-3}$ yr$^{-1}$).

\subsection{Caveats}

The number of BBH mergers matching the effective and precessing spin parameters of GW190521 is strongly affected by our choice of the spin magnitude of first-generation BHs, which is drawn from a Maxwellian distribution with $\sigma_{\chi}=0.1$. Table \ref{tab:sigma_chi} shows that changing $\sigma_{\chi}$ from 0.1 to 0.2 dramatically increases the fraction of first-generation BBHs that match GW190521's masses and spins. A choice of $\sigma_{\chi}=0.2$ would have produced $198$ first-generation BBH mergers with the same properties as GW190521, rather than just seven binaries as derived with $\sigma_{\chi}=0.1$. Hence, the merger rate density of GW190521-like systems is very sensitive to the spin distribution of first-generation BBHs: we obtain  $\mathcal{R}_{\rm GW190521}\sim{0.01}\,$Gpc$^{-3}\,$yr$^{-1}$ if $\sigma_\chi=0.01$ (no first-generation BBH mergers matching GW190521) and $\mathcal{R}_{\rm GW190521}\sim{0.47}\,$Gpc$^{-3}\,$yr$^{-1}$ if $\sigma_\chi=0.2$.

Moreover, $\mathcal{R}_{\rm GW190521}$ also depends on $\mathcal{R}_{\rm BBH}$, which in turn varies with redshift. In the LIGO-Virgo sensitivity range, this translates to a merger rate density of systems like GW190521 that ranges from $\sim0.01\,{\rm Gpc}^{-3}\,{\rm yr}^{-1}$ at $z\sim0$ up to $\sim0.04\,{\rm Gpc}^{-3}\,{\rm yr}^{-1}$ at $z\sim1$ for our fiducial model ($\sigma_{\chi}=0.1$).

We simulated a single three-body interaction for each original binary. This is a conservative approach, because each simulated BBH might undergo more than one interaction after its formation and before its ejection from the YSC. However, our simulated YSCs are relatively short lived ($\lesssim{1}$ Gyr) with a low escape velocity ($v_{\rm esc}\sim10$\,km\,s$^{-1}$), and their central density drops soon after their formation \citep{rastello2021}. Hence, it is reasonable to assume that each BBH cannot undergo a long chain of encounters. To further support our choice of a single encounter per binary, we calculated the value of the semi-major axis below which a binary can be ejected by a single--binary scattering \citep{miller2002,antonini2016}:
\begin{eqnarray}\label{eq:eq12}
    a_{\rm ej} = \frac{\xi{} \,{}m_3^2}{(m_1+m_2)^{3}}\frac{G \,{}m_1\,{} m_2}{v_{\rm esc}^{2}},
\end{eqnarray}
where $\xi=3$ \citep{quinlan1996} is a dimensionless parameter and $v_{\rm esc}$ is the escape velocity from the star cluster. We estimated that $\approx{80}\%$ of our BBHs have $a\leq{}a_{\rm ej}$ at the end of the three-body simulation. Hence, most of them are ejected from the cluster after the first encounter. This also implies that most BBHs evolve unperturbed after the simulated three-body interaction. Second-generation BBHs are therefore likely ejected from the cluster, where they can freely evolve and merge in the field. If retained, the probability of experiencing a second three-body encounter with another BH is low due to the short life span of the cluster.

In our scattering experiments we considered just triple BH interactions, without stellar components.  This assumption implies that our three-body encounters take place after all BHs, even the lightest ones, have formed in a star cluster ($t\gtrsim{}10$ Myr). While including three-body encounters between our BBHs and non-degenerate stars would make our simulations more realistic, it is unlikely that this kind of interactions drastically affect our results. Firstly, at $t\gtrsim{}10$ Myr, only stars with mass $\lesssim{15}\,$M$_{\odot}$ remain in the cluster: it is unlikely that these stars exchange with our massive BBHs. Secondly, BHs in YSCs tend to dynamically decouple from the lighter stars and to interact mainly with each other, because of their larger mass and shorter dynamical friction timescale \citep{spitzer1987,morscher2015}.

\section{Summary}\label{sec:summary}

We studied the dynamical formation of GW190521 via three-body interactions in massive YSCs. By means of direct \textit{N}-body simulations, we performed $2\times10^5$ dynamical encounters between a BBH and a single BH with mass $\geq60\,$M$_{\odot}$, above the lower edge of the PI mass gap. Our simulations include the first post-Newtonian terms (1, 2 and 2.5)  and a relativistic kick prescription for the merger remnants. 
We generate the mass, semi-major axis and orbital eccentricity of our BBHs from the population produced in the YSC simulations of \cite{dicarlo2019}. In this way, our sample includes also BHs with mass inside and above the PI gap, produced by stellar collisions in massive YSCs. We adopt a Maxwellian distribution with $\sigma_{\chi}=0.1$ to generate the magnitude of BH spins, while their direction is isotropic over the sphere \citep{bouffanais2019,bouffanais2021}.

From our simulations, we extract the first- and second-generation BBH mergers that match the main properties of GW190521 ($m_1$, $m_2$, $\chi_{\rm eff}$, $\chi_{\rm p}$, $M_{\rm rem}$, $\chi_{\rm rem}$) within the 90\% credible interval reported by \cite{AbbottGW190521astro}. About $11\%$ of our simulated BBH mergers lie inside the 90\% credible interval of the component masses of GW190521. In contrast, only 0.17\% of our simulated BBH mergers have not only the mass of the components, but also the effective and precessing spin parameters, and the final mass and spin of the BH remnant in the $90\%$ credible intervals of GW190521, as reported by \cite{AbbottGW190521}. Seven of these systems are exchanged first-generation binaries where the BH intruder replaced the secondary component of the original BBH, while five are second-generation BBHs. All the systems that match the properties of GW190521 have eccentricity $<10^{-4}$ in the LIGO--Virgo range, with the exception of two second-generation BBHs that have respectively $e\sim0.003$ ($e\sim0.4$) and $e\sim0.004$ ($e\sim0.3$) at $10$ Hz ($10^{-2}$ Hz).

All the second-generation BBHs resulting from the simulations match the observed ranges of $\chi_{\rm p}-\chi_{\rm eff}$ for GW190521, forming a separate population with non-negligible precessing spin parameter with respect to first-generation BBHs. Nevertheless, these systems are much rarer than exchanged binaries, which in turn represent almost all ($\sim{98.5\%}$) of the BBH mergers with the components in the same mass range as GW190521.

The effective and precessing spins are the most constraining parameters for GW190521-like systems in our simulations because we assumed that first-generation BHs have relatively low spins, following a Maxwellian distribution with $\sigma_{\chi}=0.1$. If we relax this assumption, many more first-generation BBHs match the main properties of GW190521 ($m_1$, $m_2$, $\chi_{\rm eff}$, $\chi_{\rm p}$, $M_{\rm rem}$, $\chi_{\rm rem}$), increasing from $0.1\%$  of all our simulated first-generation BBH mergers for $\sigma_\chi=0.1$ up to $\sim{4.2}\%$ for $\sigma_{\chi}=0.5$ (Table~\ref{tab:sigma_chi}). We do not know the exact spin distribution of massive BHs born from stellar mergers, but we can guess that high spins are possible, because the entire star collapses to BH in this scenario  \citep{costa2020}. 

If we assume relatively low spins for first-generation BHs ($\sigma_{\chi}=0.1$), the merger rate density of GW190521-like systems is $\mathcal{R}_{\rm GW190521}\sim{0.03}\,$Gpc$^{-3}\,$yr$^{-1}$, within the 90\% credible interval derived by \cite{AbbottO3IMBH} but rather on the low side. Our estimate of the merger rate density  is very sensitive to the spin distribution of first-generation BBH mergers: we obtain $\mathcal{R}_{\rm GW190521}\sim{0.01}\,$Gpc$^{-3}\,$yr$^{-1}$ if $\sigma_\chi=0.01$ (no first-generation BBH mergers matching GW190521) and $\mathcal{R}_{\rm GW190521}\sim{0.46}\,$Gpc$^{-3}\,$yr$^{-1}$ if $\sigma_\chi=0.2$.
Our results imply that GW190521, if it was born in a massive YSC, is either a first-generation BBH resulting from an exchange with a massive intruder ($\geq60\,$M$_{\odot}$) or a second-generation BBH merger triggered by a resonant three-body encounter.

\section*{Acknowledgements}

We thank the anonymous Referee for their useful comments which helped us improve our work. MD acknowledges financial support from Cariparo foundation under grant 55440. MM, UNDC, YB, SR, FS and AB acknowledge financial support by the European Research Council for the ERC Consolidator grant DEMOBLACK, under contract no. 770017.  MAS acknowledges financial support from the Alexander von Humboldt Stiftung under the research program "Black Holes at all the scales" , the Volkswagen Foundation Trilateral Partnership project No. I/97778 “Dynamical Mechanisms of Accretion in Galactic Nuclei”, the Deutsche Forschungsgemeinschaft (DFG, German Research Foundation) – Project-ID 138713538 – SFB 881 (“The Milky Way System”), and the COST Action CA16104. We acknowledge the CINECA-INFN agreement for the availability of high performance computing resources and support. We also thank Roberto Capuzzo Dolcetta, Pauline Chassonnery and Seppo Mikkola for making the {\sc arwv} code available to us.

\section*{Data availability}
The data underlying this article will be shared on reasonable request to the corresponding authors.



\bibliographystyle{mnras}
\bibliography{bibliography} % if your bibtex file is called example.bib

%%%%%%%%%%%%%%%%%%%%%%%%%%%%%%%%%%%%%%%%%%%%%%%%%%

%%%%%%%%%%%%%%%%% APPENDICES %%%%%%%%%%%%%%%%%%%%%

%\appendix


%%%%%%%%%%%%%%%%%%%%%%%%%%%%%%%%%%%%%%%%%%%%%%%%%%

% Don't change these lines
\bsp	% typesetting comment
\label{lastpage}
\end{document}